\documentclass[pra,showpacs,twocolumn,amsmath]{revtex4}
\usepackage{amssymb}
\usepackage{amsmath}
\usepackage{epsfig}
\usepackage{amsfonts}
\usepackage{color}
\usepackage{float}
\usepackage{latexsym}
\usepackage{mathrsfs}
\usepackage{feynmf}
\def\QED{\leavevmode\unskip\penalty9999 \hbox{}\nobreak\hfill
     \quad\hbox{\leavevmode  \hbox to.77778em{%
               \hfil\vrule   \vbox to.675em%
               {\hrule width.6em\vfil\hrule}\vrule\hfil}}
     \par\vskip3pt}
\def\qed{\leavevmode\unskip\penalty9999 \hbox{}\nobreak\hfill
     \quad\hbox{\leavevmode  \hbox to.77778em{%
               \hfil\vrule   \vbox to.675em%
               {\hrule width.6em\vfil\hrule}\vrule\hfil}}
\par\vskip3pt}
\def\ibb #1{\leavevmode\hbox{\kern.3em\vrule
     height 1.5ex depth -.1ex width .4pt\kern-.3em\rm#1}}

\newcommand{\be}{\begin{equation}}
\newcommand{\ee}{\end{equation}}
\newcommand{\ba}{\begin{array}}
\newcommand{\ea}{\end{array}}
\newcommand{\bqa}{\begin{eqnarray}}
\newcommand{\eqa}{\end{eqnarray}}

\begin{document}
\title{Multipartite Quantum Correlations in Open Quantum Systems}
\author{Zhi-Hao Ma$^{1,2}$}
\affiliation{$^1$ Department of Mathematics, Shanghai Jiaotong
University, Shanghai, 200240,  China \\
$^2$Department of Physics and Astronomy, University College
London, Gower St., WC1E 6BT London, United Kingdom}
\author{Zhi-Hua Chen$^{3}$}
\affiliation{$^3$Department of Science, Zhijiang college, Zhejiang
University of technology, Hangzhou, 310024, P.R.China}
\author{Felipe Fernandes Fanchini$^{4}$}
\affiliation{$^4$Departamento de F\'{\i}sica, Faculdade de Ci\^encias, Universidade Estadual Paulista, Bauru, SP, CEP 17033-360, Brazil, corresponding author, fanchini@fc.unesp.br}

\begin{abstract}
In this paper we present a measure of quantum correlation for a multipartite system, defined as the sum of the correlations for all possible partitions. Our measure can be defined for quantum discord (QD), geometric quantum discord (GQD), or even by the entanglement of formation (EOF). For tripartite pure states, we show that the multipartite measure for the QD and the EOF are equivalent, which gives a way to compare the distribution and the robustness of these correlations in open quantum systems.
We study dissipative dynamics for two distinct families of entanglement: a W state and a GHZ state. We show that, while for the W state the QD is more robust than the entanglement, for the GHZ state this is not true. It turns out that the initial genuine multipartite entanglement present in the GHZ state makes the EOF more robust than the QD.
\end{abstract}


\maketitle

\section{Introduction}

 Entanglement in composite quantum systems leads to many puzzling paradoxes in quantum theory \cite{Peres,Horodecki09,Guhne09,Fuchs11}.
The importance of entanglement is recognized unanimously but it is well known that even separable quantum states possess correlations that cannot be simulated by classical systems; e.g., the nonclassical correlations captured by quantum discord (QD) \cite{qd}. For bipartite state, many efforts were done to detect and quantify QD \cite{Girolami,Yu} and find the connection with entanglement \cite{law, Piani}. 
 Moreover, much attention is being devoted to the quantify of quantum correlation in multipartite systems \cite{rulli, other}. Such measures help us to understand the distribution of quantum correlations and provide a way of studying dissipative dynamics in many-part quantum systems. 
 
 Although it is claimed that the QD is more robust than the entanglement in open quantum systems, little concrete evidence has been published to corroborate this claim where multipartite systems are concerned.
 Actually, for two qubits there is a good deal of evidence that it is true \cite{robust}, but can we extend this rule to multipartite quantum systems?
For three qubits, for example, there exist two distinct families of entanglement \cite{family}; so, how robust, then, is the quantum discord for members of each family?  Could entanglement be more robust than discord, against noise arising from the environment, in multipartite systems? There are very few studies on these questions in the literature \cite{openqd} and that is the focus of this article.

To investigate the robustness of the quantum correlations in open quantum systems for tripartite states, we define a multipartite measure of quantum correlation slightly different from that employed by Rulli and Sarandy \cite{rulli}. Those the authors define a measure of the global multipartite quantum discord as the maximum of the quantum correlations that exist among all possible bipartitions. Here, to attain an average measure, we define global QD as the sum of correlations for all possible bipartitions. Despite being only subtly different, our measure now accounts for how the quantum correlation is distributed in the tripartite system and certainly gives a better insight into the robustness of these correlations in open quantum systems.

This paper is organized as follows: in section II we present the formal definition of the entanglement of formation, the usual quantum discord, and the geometric measure of quantum discord. In section III we define our multipartite measure of quantum correlations (MMQC) based on the sum of correlations for all possible bipartitions. In section IV we present an analytical solution to the MMQC for a three-qubit pure state and we show that for a general tripartite pure state the MMQC based on the usual quantum discord and on the entanglement of formation are equivalent. In section V we extend our analysis to a tripartite mixed state considering the dissipative dynamics of three qubits and in section VI we summarize our results.
\section{Quantum Correlations}
In this manuscript we consider three well-known measures of quantum correlations: entanglement of formation, usual quantum discord, and the geometric measure of quantum discord. Below, we present the formal definition of each one of these measurements.
\subsection{Entanglement of Formation}
The entanglement of formation is a measure of entanglement defined more than fifteen years ago by Bennett, DiVincenzo, Smolin, and Wootters \cite{eof}. Although very different from QD, it is connected with the latter a monogamic relation \cite{koashi,law} and has a nice operational interpretation. It is defined as follows. Given a bipartite system $A$ and $B$, consider all possible pure-state decompositions of the density matrix $\rho_{AB}$; that is, all ensembles of states $|\Psi_i\rangle$ with probability $p_i$ such that $\rho_{AB}=\sum_i p_i |\Psi_i\rangle\langle\Psi_i|$. For each pure state, the entanglement is defined as the von Neumann entropy of either of the two subsystems $A$ and $B$, such that $E(\Psi)=S_A=S_B$, where $S_A:=S(\rho_A)=-{\rm{Tr}}(\rho_A\log\rho_A)$, $\rho_A$ being the partial trace over $B$, and there is an analogous expression for $S_B$. The EOF for a mixed state is the average entanglement of the pure states minimized over all possible decompositions, i. e.,
\begin{equation}
E(\rho_{AB})=\min \sum_i p_i E(\Psi_i).
\end{equation}
Whereas this is very hard to calculate for a general bipartite system, for two qubits there is an analytical solution given in the seminal William K. Wootters paper \cite{conc}.

\subsection{Usual Quantum Discord}
Quantum discord is a well known measure of quantum correlation defined by Ollivier and Zurek \cite{qd} about ten years ago. It is defined as
\begin{equation}
\delta_{AB}^\leftarrow=I_{AB}-J_{AB}^\leftarrow
\end{equation}
where $I_{AB}=S_A + S_B - S_{AB}$ is the mutual information and $J_{AB}^\leftarrow$ is the classical correlation \cite{cc}. Explicitly,
\begin{equation}
J_{AB}^\leftarrow=\max_{\{\Pi_k\}}\left[ S_A - \sum_k p_k S(\rho_{A|k})\right],
\end{equation}
where $\rho_{A|k} = {\rm{Tr}}_B\left(\Pi_k \rho_{AB} \Pi_k\right)/{\rm {Tr}}_{AB}\left(\Pi_k \rho_{AB} \Pi_k\right)$ is the locally post-measured state after obtaining the outcome $k$ in $B$ with probability $p_k$. The quantum discord measures the amount of the mutual information that is not accessible locally \cite{demon, LII} and generally is not symmetric, i.e. $\delta_{AB}^\leftarrow\ne\delta_{BA}^\leftarrow$.

\subsection{Geometric Measure of Quantum Discord}

Intuitively,  quantum discord can be viewed as a measure of the minimum loss of correlation due to measurements in the sense of quantum
mutual information. A state with zero discord, i.e., $\delta_{AB}^\leftarrow=0$, is a state whose information is not disturbed by local measurements; it is called as a classical-quantum (CQ) state. A CQ state  is of the form \cite{qd}
\begin{equation}
\chi=\sum\limits_{k=1}^{m}p_k|k\rangle\langle k|\otimes\rho_k^{B}
\end{equation}
where $\{p_{k}\}$ is a probability distribution, $|k\rangle$ an arbitrary orthonormal basis in subsystem $A$ and $\rho_k^{B}$ a
set of arbitrary density matrices in subsystem $B$.

Denoting the set of all CQ states on $H_{A}\otimes H_{B}$  as $\Omega_{0}$, then it is natural to think that, the further a state $\rho$ is from $\Omega_{0}$, the higher is its quantum discord.
Indeed, we can use the distance from $\rho$ to the nearest state in $\Omega_{0}$ as a measure of discord for the state, and this is the idea behind the geometric measure of quantum discord (GQD) \cite{dakic}.
Thus, the GQD was introduced as:
\begin{equation}
D(\rho)=\mathrm{min}_{\chi\in\Omega_0}||\rho-\chi||^2,
\end{equation}
where $\Omega_0$ denotes the set of zero-discord  states and $||X-Y||^2=\mathrm{tr}(X-Y)^2$ stands for the squared Hilbert-Schmidt norm. Note that the maximum value reached by the GQD is $\frac{1}{2}$ for two-qubit states, so it is appropriate to consider $2D$ as a measure of GQD hereafter, in order to compare it with other measures of correlation \cite{adesso}.

Interestingly, an explicit expression for GQD in a general two-qubit state can be written \cite{dakic}. In the Bloch representation, any two-qubit state $\rho$ can be represented as follows:
\begin{equation}
\rho=\frac{1}{4}(\mathbb{I}\otimes \mathbb{I}+\sum_{i=1}^3x_i\sigma^i\otimes \mathbb{I}+\sum_{i=1}^3y_i\mathbb{I}\otimes\sigma^i+\sum_{i,j=1}^3t_{ij}\sigma^i\otimes\sigma^j),
\end{equation}
where $\mathbb{I}$ is the identity matrix, $\sigma^i (i=1,2,3)$ are the three Pauli matrices, $x_i=\mathrm{tr}(\sigma^i\otimes\mathbb{I})\rho$ and $y_i=\mathrm{tr}(\mathbb{I}\otimes\sigma^i)\rho$ are the components of the local Bloch vectors $\vec{x}$ and $\vec{y}$, respectively, and $t_{ij}=\mathrm{tr}(\sigma^i\otimes\sigma^j)\rho$ are components of the correlation matrix $T$.  Then the GQD of $\rho$ is given by
\begin{equation}
D(\rho)=\frac{1}{4}(||\vec{x}||^2+||T||^2-\lambda_{\mathrm{max}}),\label{D}
\end{equation}
$\lambda_{\mathrm{max}}$ being the largest eigenvalue of the matrix $K=\vec{x}\vec{x}^t+TT^t$ and $||T||^2=\mathrm{tr}(TT^t)$. The superscript $t$ denotes the transpose of vectors or matrices. Furthermore, it is important to mention that an analytical solution for a bipartite system of dimension $2\times N$ has been given in \cite{agd}.

\section{Multipartite Measure of Quantum Correlations}

{\bf Definition 1.} For an arbitrary $N$-partite state $\hat{\rho}_{1 \cdots N}$, the {multipartite measure of quantum correlation} $\mathcal{Q}\left( \hat{\rho}_{1 \cdots N} \right)$ is defined as follows:

Let $\rho$ be an $N$-partite state, and $\mu$ and $\nu$ be any subsets among all possible partitions. The  multipartite measure of quantum correlations (MMQC) is defined as the sum of the quantum correlations for all possible bipartitions,
\begin{eqnarray}
{\mathcal Q}\left( \hat{\rho}_{1 \cdots N} \right) = \sum_{\mu\ne\nu=1}^N M_{\mu(\nu)}\label{smmqc}
\end{eqnarray}
where $M_{\mu(\nu)}$ is a measure of quantum correlation that can be given by the geometric quantum discord, the usual quantum discord or the entanglement of formation. Here, the subset between $(\cdot)$ is the measured one in the case of GQD or QD and can be ignored for the entanglement of formation. As we can see from the definition given in Eq. (\ref{smmqc}), our measure is symmetrical and, more importantly, it gives zero if the state has just classical-classical correlations.

To elucidate the MMQC defined above, let us consider the case of a tripartite state. Explicitly,
\begin{eqnarray}
{\mathcal Q}\left( \hat{\rho}_{ABC} \right) &=& M_{A(B)} + M_{B(A)} + M_{A(C)} + M_{C(A)}\nonumber\\
 &+&\!\! M_{B(C)} + M_{C(B)} + M_{A(BC)} + M_{BC(A)}\nonumber\\
 &+&\!\! M_{B(AC)} + M_{AC(B)} + M_{C(AB)} + M_{AB(C)}.\;\;\;\;\;\;\label{mmqc}
\end{eqnarray}

\section{MMQC analytical solution for three qubits pure state}
Our starting point is an analytical solution to the MMQC for three-qubit pure states. For the case of geometric measure of quantum discord, an analytical solution can be found with the help of the result presented in \cite{agd}. Actually, from that result, an analytical solution can be obtained for GQD even for three-qubit mixed states. The analytical solution for three qubits can also be obtained for the usual EOF. With the help of Wootters concurrence \cite{conc}, each entanglement measurement involving this sets of one subsystem (e. g. $E_{B(C)}$) can be obtained trivially by calculating the eigenvalues of the density matrix. Furthermore, we note that the entanglement measure involving a set of subsystem and a set of two (e.g. $E_{A(BC)}$), is given by the von Neumann entropy of one of the partitions. For example, $E_{A(BC)}=S_A=S_{BC}$.

Finally, to calculate the MMQC for the usual quantum discord, we note the result given in \cite{LII}, where the authors show that the sum of the quantum discord for all possible bipartitions, involving sets of one subsystem, is equal to the sum of the entanglement of formation for all possible bipartitions,
\begin{eqnarray}
&&E_{A(B)} + E_{B(A)} + E_{A(C)} + E_{C(A)} + E_{B(C)} + E_{C(B)} = \nonumber\\
&&\delta_{A(B)}^\leftarrow + \delta_{B(A)}^\leftarrow  + \delta_{A(C)}^\leftarrow  + \delta_{C(A)}^\leftarrow  + \delta_{B(C)}^\leftarrow  + \delta_{C(B)}^\leftarrow.\label{sum}
\end{eqnarray}
Moreover, since we are considering pure states, the quantum discord measure involving a set of one and a set of two subsystems is given by the entropy of one of the partitions, exactly as the entanglement of formation. For example, $E_{A(BC)}=\delta_{A(BC)}^\leftarrow=\delta_{BC(A)}^\leftarrow=S_A=S_{BC}$.

Thus, for general tripartite pure states, the MMQC defined in Eq. (\ref{mmqc}) is identical for the QD and for the EOF which results in an analytical solution for the MMQC for three-qubit pure states, for the QD as well. To confirm this, we note the result given in Eq. (\ref{sum}), which is valid irrespective of the system dimension of the subsystems. On the other hand, for general tripartite states (not three qubits), an analytical solution does not exist, either for the EOF or for the QD.

\section{MMQC for three-qubit mixed states}
To calculate the MMQC for three-qubit mixed states, we limit ourselves to studying a rank-2 density matrix. In this case, as we show below, a simple strategy can be used. Since we are considering three qubits, to calculate the MMQC for sets of one subsystem ($E_{A(B)}$, $E_{A(C)}$, $\delta_{A(B)}^\leftarrow$, $\delta_{A(C)}^\leftarrow$, etc.) is trivial. In this case, the EOF can be calculated analytically by means of concurrence, and the quantum discord can be calculated numerically by using positive-operator valued measurements (POVM) \cite{gian}. The question is: how can we calculate the MMQC for the terms that involves a set of one subsystem and a set of two, i.e. $E_{A(BC)}$, $E_{B(AC)}$, $\delta_{A(BC)}^\leftarrow$, $\delta_{B(AC)}^\leftarrow$, etc.? Here, in contrast to the case of pure states, the von Neumann entropy does not give the right answer. The answer to this question is given by the monogamic relation between entanglement of formation and quantum discord \cite{koashi,law}.

Since we are considering a rank-2 density matrix, an extra subsystem $E$ that purifies $ABC$ is a two-level system. Thus, the idea is to calculate the quadripartite pure state $\rho_{ABCE}$ and use the monogamic relation. The monogamic relation implies that the EOF between two partitions is connected with the QD between one of the partition and the third one that purifies the pair. To elucidate this further, we present in detail the strategy used to calculate the QD and the EOF for the rank-2 tripartite mixed state $\rho_{ABC}$. Here, we show the way to calculate just the terms $E_{A(BC)}$ and $\delta_{A(BC)}^\leftarrow$, but the strategy is analogous for all the other terms involving a set of one subsystem and a set of two given in Eq. (\ref{mmqc}).
For the EOF, the monogamic relation says that
\begin{equation}
E_{A(BC)}=\delta_{A(E)}^\leftarrow + S_{A|E},
\end{equation}
where $A(E)$ involves two qubits, since $ABC$ is of rank 2. In this case, we can calculate $\delta_{A(E)}^\leftarrow$ numerically. For the quantum discord, on the other hand, the result can be reached by
\begin{equation}
\delta_{A(BC)}^\leftarrow=E_{A(E)}+ S_{A|E},
\end{equation}
where $E_{A(E)}$ can be calculated by means of the concurrence. Moreover, $\delta_{BC(A)}^\leftarrow$ can be calculated numerically, since the measurements are just over one qubit, in this case, subsystem $A$. Thus, following the recipe above, it is straightforward to calculate the quantum correlations for all terms of Eq. (\ref{mmqc}).

\section{MMQC in open quantum systems}
To calculate the MMQC in open quantum systems, we study two special situations: firstly, a three-qubit W-state subjected to independent amplitude-damping channel and secondly, the GHZ state, with independent phase damping. The reason for this specific choice is that, for the whole of the dissipative process, we have a rank-2 density matrix \cite{thiago}. In this case, we can use the strategy explained above to calculate the MMQC for the usual QD and the EOF.

The dissipative dynamics can be calculated straightforwardly by means of Kraus operators \cite{nielsen}. Since we consider independent environments for each qubit, given an initial state for three qubits $\rho(0)$, its evolution can be written as
\begin{equation}
\rho(t)=\sum_{\alpha,\beta,\gamma}E_{\alpha,\beta,\gamma}\rho(0)E^\dagger_{\alpha,\beta,\gamma},
\end{equation}
where the so-called Kraus operators $E_{\alpha,\beta,\gamma}\equiv E_{\alpha}\otimes E_{\beta}\otimes E_{\gamma}$ satisfy $\sum_{\alpha,\beta,\gamma}E^\dagger_{\alpha,\beta,\gamma}E_{\alpha,\beta,\gamma}=\mathbb{I}$ for all $t$. The operators $E_{\{\alpha\}}$ describe the one-qubit quantum channel effects.
We first consider a W-state subjected to independent amplitude damping. This damping describes the exchange of energy between the system and the environment and is described by the Kraus operators $E_0=\sqrt{p}(\sigma_x+i\sigma_y)/2$ and $E_1={\rm diag}(1,\sqrt{1-p})$, where $p=1-e^{-\Gamma t}$, $\Gamma$ denoting the decay rate, and $\sigma_x$ and $\sigma_y$ are Pauli matrices.
\begin{figure}[htbp]
\begin{center}
\includegraphics[width=.45\textwidth]{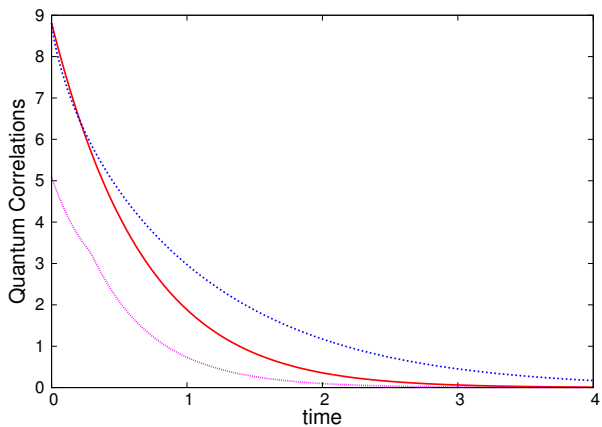} {}
\end{center}
\caption{(Color Online) the red curve (solid) shows the EOF dynamics, while the blue curve (dotted) shows QD and the cyan curve (traced) the GQD: initial W state subjected to independent amplitude-damping channels.}\label{fig1}
\end{figure}
 In Fig. (\ref{fig1}) we show the dissipative dynamics of the MMQC for an initial state given by $|W\rangle =(|100\rangle + |010\rangle + |001\rangle)/\sqrt{3}$. We see that in this situation the QD is actually more robust than EOF and GQD. For a very short time, the EOF resists but it decays fast while QD maintains it robustness. This result corroborates the idea that the QD is more robust than the EOF in open quantum systems. In the second, and more important case, we consider an initial GHZ state subjected to phase damping.  The dephasing channel induces a loss of quantum coherence without any energy exchange. In this case the Kraus operators are given by $E_0={\rm diag}(1,\sqrt{1-p})$ and $E_1={\rm diag }(1,\sqrt{p})$.
 \begin{figure}[htbp]
\begin{center}
\includegraphics[width=.45\textwidth]{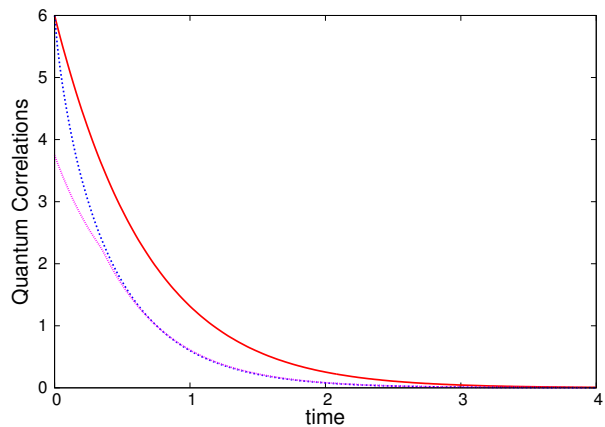} {}
\end{center}
\caption{(Color Online) the red curve (solid) shows the EOF dynamics, while the blue curve (dotted) shows QD, and the cyan curve (traced) represents the GQD. All of them for the initial GHZ state subjected to independent phase damping channels.}\label{fig2}
\end{figure}
 In Fig. (\ref{fig2}) we show the dissipative dynamics of the MMQC for an initial state given by $|GHZ\rangle=(|000\rangle + |111\rangle)/\sqrt{2}$. Here, we see a very interesting result. Contrary to what is claimed in the literature, the EOF is in fact more robust than the QD for this kind of initial condition and dissipative channel. EOF is sustained for a longer time than QD.

This peculiar situation occurs because of the conservative relation between the EOF and the QD \cite{law}. For the W state, the QD between the parts of the system is sustained and, consequently, these subsystems need to have more entanglement with the environment to preserve the conservative relation. Indeed, it is impossible to create or destroy some of the EOF (QD) without destroying or creating the same amount of QD (EOF). For the W state subjected to the amplitude-damping channel, each subsystem becomes more entangled with the environment but creates little discord. So, the QD is sustained in the system to maintain the conservative relation. On the other hand, for the GHZ state we have a genuine multipartite quantum correlation, since the MMQC for the EOF or the QD is given by:
\begin{eqnarray}
{\mathcal Q}\left( \hat{\rho}_{ABC} \right) &=& M_{A(BC)} + M_{BC(A)} + M_{B(AC)} \nonumber\\
 &+& M_{AC(B)} + M_{C(AB)} + M_{AB(C)}.\label{mmqc2}
\end{eqnarray}
To calculate Eq. (\ref{mmqc2}) for the quantum discord, we use, as explained above, the monogamic relation between the EOF and the QD. In this case, each one of the six terms is given by
\begin{eqnarray}
\delta^\leftarrow_{A(BC)}=\delta^\leftarrow_{BC(A)}&=& E_{A(E)}+S_{A|E},\nonumber\\
\delta^\leftarrow_{B(AC)}=\delta^\leftarrow_{AC(B)}&=& E_{B(E)}+S_{B|E},\nonumber\\
\delta^\leftarrow_{C(AB)}=\delta^\leftarrow_{AB(C)}&=& E_{C(E)}+S_{C|E}.\label{mmqc3}
\end{eqnarray}
However, there is no entanglement between each subsystem and the environment, i.e.  $E_{A(E)}=E_{B(E)}=E_{C(E)}=0$, since the MMQC for the QD can be calculated analytically by means of the conditional entropy. Indeed,
\begin{eqnarray}
{\mathcal Q}_{QD}\left( \hat{\rho}_{ABC} \right) &=& 2(S_{A|E} + S_{B|E} + S_{C|E}).
\end{eqnarray}
For the EOF, the situation is a little different. In this case, each one of the six terms is given by
\begin{eqnarray}
E_{A(BC)}=E_{BC(A)}&=& \delta^\leftarrow_{A(E)}+S_{A|E},\nonumber\\
E_{B(AC)}=E_{AC(B)}&=& \delta^\leftarrow_{B(E)}+S_{B|E},\nonumber\\
E_{C(AB)}=E_{AB(C)}&=& \delta^\leftarrow_{C(E)}+S_{C|E}.\label{mmqc4}
\end{eqnarray}
The crucial difference between Eq. (\ref{mmqc3}) and Eq. (\ref{mmqc4}) is that, while each part of the system does not become entangled with the environment, it does create QD with it . Hence, $\delta^\leftarrow_{A(E)}=\delta^\leftarrow_{B(E)}=\delta^\leftarrow_{C(E)}\ne0$ and, consequently,
\begin{eqnarray}
{\mathcal Q}_{EOF}\left( \hat{\rho}_{ABC} \right)\!\!&=&\!\! {\mathcal Q}_{QD}\left( \hat{\rho}_{ABC} \right) + 2(\delta^\leftarrow_{A(E)} + \delta^\leftarrow_{B(E)} + \delta^\leftarrow_{C(E)}).\nonumber
\end{eqnarray}
This is an important result and a direct consequence of the conservative relation between the EOF and the QD. For the GHZ state subjected to phase damping, there is no entanglement between each subsystem and the environment but there is QD. Thus, the QD that is created with the environment needs to be compensated by the entanglement retained in the system, making the EOF more robust than the QD in this particular situation. It must be emphasized that this is a direct consequence of the GHZ being a genuine multipartite entangled state, since this means that the EOF between sets of one party ($E_{A(B)}$, $E_{A(E)}$, $E_{B(C)}$, etc.) is always zero during the dissipative dynamics.

\section{Conclusion}
In this paper, we have presented an alternative measure of multipartite quantum correlations. Our measure gives a novel and intuitive means of comparing the robustness of entanglement and discord in multipartite systems, against the detrimental interaction with the environment. We analyze two distinct initial conditions, which represent different families of multipartite entanglement.  We show that the robustness of the EOF and QD depends on the family of entanglement to which the initial state pertains, putting into question the superior robustness of QD in open quantum systems.
Actually, for a three-qubit W state, QD proves to be more robust, but the same can not be said of the GHZ state. We show that this behavior is related to the way the multipartite quantum state is quantum correlated with the environment. For the GHZ state subjected to independent phase-damping channels, contrary to the QD, the individual qubits do not get entangled with the environment, and the initial multipartite entanglement is thus preserved for a longer time than the QD. We believe that the discussion presented here may contribute further to the understanding of the distribution of entanglement and discord in open quantum systems.

\vskip 0.1 in {\noindent\bf Acknowledgment.} This work was in part supported by NSFC(10901103) and by FAPESP and CNPq through the National Institute for Science and Technology of Quantum Information (INCT-IQ).

\end{document}